\documentclass[12pt]{iopart}
\usepackage{iopams,amssymb}
\usepackage{xcolor,graphicx}

\newcommand{\vF}{v_{\mathrm{F}}}
\newcommand{\gap}{\Delta}

\newcommand{\trans}{_{\mathrm{tr}}}
\newcommand{\TE}{^{\mathrm{TE}}}
\newcommand{\TM}{^{\mathrm{TM}}}

\newcommand{\atanh}[1]{\,\mathrm{arctanh}\left(#1\right)\,}
\newcommand{\atan}[1]{\,\mathrm{arctan}\left(#1\right)\,}

\newcommand{\ii}{\mathrm{i}}
\newcommand{\ee}[1]{\mathrm{e}^{#1}}

\newcommand{\transferMatrix}{\mathcal{T}}

\begin{document}

\title{TE resonances in graphene-dielectric structures}

\author{J. F. M. Werra$^{1}$, F. Intravaia$^{2}$ and K Busch$^{1,2}$}
\address{$^{1}$ Humboldt-Universit\"{a}t zu Berlin, Institut f\"{u}r Physik, 
                AG Theoretische Optik \& Photonik, 12489 Berlin, Germany}
\address{$^{2}$ Max Born Institute, 12489 Berlin, Germany}
\ead{jwerra@physik.hu-berlin.de}
\begin{abstract}
We investigate the dispersion relations of TE resonances in different 
graphene-dielectric structures. Previous work 
has shown that when a graphene layer is brought into contact with a dielectric 
material, a gap can appear in its electric band structure. 
This allows for the formation of TE-plasmons with unusual dispersion relations. 
In addition, if the dielectric has a finite thickness, graphene strongly modifies the behavior of the waveguiding modes
by introducing 
dissipation above a well-defined cutoff frequency thus providing the possibility 
of mode filtering. This cutoff and the properties of TE-plasmons are closely related 
to the pair-creation threshold of graphene thus representing quantum mechanical effects 
that manifest themselves in the electromagnetic response.
\end{abstract}
\maketitle

\section{Introduction}

The physics of graphene is often described in terms of a (2 + 1)-dimensional 
Dirac field where the speed of light is replaced by the Fermi velocity, $\vF$. 
For pristine graphene in vacuum the effective Dirac field is massless and the 
resulting energy band structure near the Fermi level is characterized by the 
very well know linear dispersion relation (Dirac cone) associated with the 
K-points at the corners of the Brilloin zone
\cite{McCann_2012}. 
Previous work has shown, however, that in certain circumstances graphene's 
band structure can feature an electronic band gap $2\gap$ ($\gap\approx5-50\,
\mathrm{meV}$ \cite{Giovannetti_2007,Zhou_2007,Jung_2015}, see Fig.~
\ref{fig:grapheneSlabsSketch4}) which can be modeled as an effective mass in 
the Dirac model. This effective mass can be related to the occurrence of one 
or several physical phenomena, such as the presence of impurities, the contact 
with another medium (i.e., substrate), strain, etc. 
The modified electronic band structure affects the optical response of graphene
and, specifically, also its peculiar plasmonic resonances. These interesting 
features are attracting growing attention as witnessed by the vast number of 
recent publications addressing their properties (for instance, see
Refs.~\cite{Christensen_2011,Christensen_2014,Alkorre_2015,Jadidi_2015,Kumar_2015,Cai_2015}). 
In contrast to other two-dimensional electron gases that only exhibit ``ordinary'' 
TM-polarized plasmon resonances, graphene also exhibits a resonant response to 
TE-polarized light 
\cite{Mikhailov_2007}. 
When utilizing a gapless model, it turns out that while TE-plasmons 
remain when graphene is embedded into a dielectric material, they disappear when 
graphene is sandwiched between two different dielectrics~
\cite{Kotov_2013,Stauber_2014}. 
However, when graphene is deposited on a substrate (e.g., hexagon boron-nitride 
or silicon carbite), the offset in lattice spacing between the two materials can
lead to the opening of a relatively wide band gap~
\cite{Giovannetti_2007,Zhou_2007,Jung_2015,Gusynin_2007_tightBinding}, 
rendering a gapless description inadequate.
In this work, we investigate the existence of TE plasmons for certain 
prototypical graphene-dielectric configurations, introducing the gap as 
phenomenological parameter on the order of tens of meV (see, for instance, the ab-initio study
~\cite{Jung_2015}). We demonstrate how the appearance of a band gap 
and the pair-creation threshold affect the dispersion relations 
of the TE resonances, 
conferring distinct quantum feature upon them. 

\section{Resonances in graphene-dielectric systems}

\begin{figure}[t!]
\centering
  \includegraphics{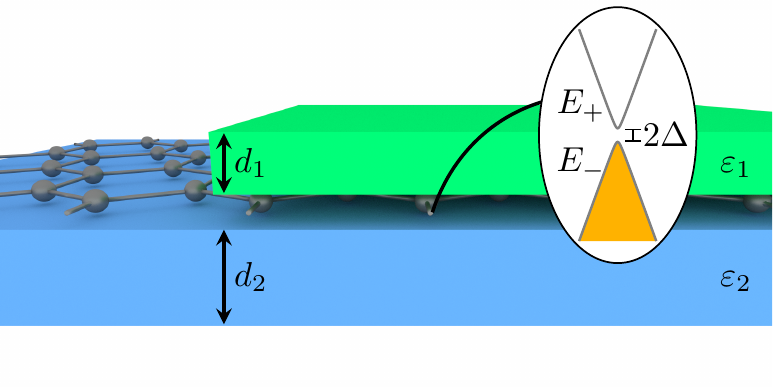}
\caption{Schematics of the graphene-dielectric structure considered in this work. 
         The inset depicts the electronic band structure of graphene in the vicinity
         of the Dirac cone $E_{\pm}=\pm\sqrt{\gap^2+\hbar^2\vF^2k^2}$. Here, the
         in-plane wavevector $k$ is measured from the K-points of the Brillouin 
         zone and $\vF$ denotes the Fermi velocity.}
\label{fig:grapheneSlabsSketch4}
\end{figure}

In the following, we employ dimensionless units such that velocities are expressed
in units of the vacuum speed of light $c$ and frequencies are measured in units of
the band gap $2\Delta / \hbar$. This amounts to the replacements $\vF/c\to\vF$ and
$\hbar\omega/(2\gap)\to\omega$, respectively, and the in-plane wavevector 
changes accordingly $\hbar c k/(2\gap)\to k$.
Following the quantum-field theoretical description of graphene discussed in earlier 
works
\cite{Fialkovsky_2011,Chaichian_2012,Bordag_2014,Bordag_2015},
we relate its optical properties to two components of the polarization tensor
\begin{eqnarray}
\Pi_{00} &=
\alpha\frac{k^2}{\vF^2k^2-\omega^2}\Phi(\omega,k)\,,\quad
 \Pi\trans &= \alpha\frac{(\vF^2+1)k^2-2\omega^2}{\vF^2k^2-\omega^2}\Phi(\omega,k),
\end{eqnarray}
where we have introduced $\Phi(\omega,k)\equiv\Phi(y)=2\left[1-\frac{y^2+1}{y^2}y\atanh{y}\right]$
with $y=\sqrt{\omega^2-\vF^2k^2}$ and $\alpha = 137^{-1}$ denotes the fine structure constant.
Using these quantities, the electromagnetic reflection and transmission coefficients 
for a single and freely suspended graphene layer are given by~
\cite{Fialkovsky_2011,Chaichian_2012}
\begin{eqnarray}
\label{eq:rTMgraphene}
 r_g\TM&=\frac{\kappa_0\Pi_{00}}{k^2\Pi\trans-\kappa_0^2\Pi_{00}+2k^2\kappa_0}\,,\qquad &t_g\TM=1-r_g\TM\,,\\
\label{eq:rTEgraphene}
 r_g\TE&=\frac{k^2\Pi\trans-\kappa_0^2\Pi_{00}}{\kappa_0\Pi_{00}+2k^2}\,,\qquad &t_g\TE=1+r_g\TE\,.
\end{eqnarray}
Here, we have defined $\kappa_0=\sqrt{k^2-\omega^2}\equiv-\ii k_{0}^{\perp}$, which 
is related to the vacuum wavevector component $k_0^\perp$ orthogonal to the plane
of the graphene layer. When a single graphene layer is embedded between two slabs made 
of different dielectric materials and/or different thicknesses (cf. Fig.~\ref{fig:grapheneSlabsSketch4}), 
the reflection and transmission coefficients can be found using the transfer
matrix technique~\cite{Klimchitskaya_2014,Klimchitskaya_2015} 
\begin{eqnarray}
 \transferMatrix&=\transferMatrix_{2}\cdot\transferMatrix_{\mathrm{graphene}}\cdot\transferMatrix_{1} \, ,\qquad
\mathrm{where}\,\,\transferMatrix_i=\frac{1}{t_i}\left(\begin{array}{@{}cc@{}} t_i^2-r_i^2
& r_i \\ -r_i
& 1 \end{array}\right)\,.
\end{eqnarray}
The expression of $\transferMatrix_i$ is characteristic for a homogeneous dielectric 
layer~
\cite{Saleh_Teich_engl} 
where $r_i$ and $t_i$, respectively, represent the layer's reflection and transmission 
coefficients. From a straightforward calculation for illumination from the side of 
layer 1, we obtain that the scattering properties of the multilayer structure are 
described by
\begin{eqnarray}
 r&=r_1+t_1^2\frac{r_g+r_2\left(t_g^2-r_g^2\right)}{(r_2r_g-1)(r_1r_g-1)-r_1r_2t_g^2}\,,\\
 t&=\frac{t_1t_gt_2}{(r_2r_g-1)(r_1r_g-1)-r_1r_2t_g^2}\,.
\end{eqnarray}
While for the reflection and transmission coefficients of graphene, $r_g$ an $t_g$, we can 
utilze Eqs.~\eref{eq:rTMgraphene} and \eref{eq:rTEgraphene}, the dielectric layers' scattering 
coefficients, $r_i$ and $t_i$, depend on the thickness $d_{i}$ and on the material permittivity 
$\varepsilon_{i}$. For the case of TE polarized radiation and considering the total structure with
vacuum on both sides we explicitly
obtain 
\begin{eqnarray}
\label{eq:rtTEslab}
 r_i&=\frac{\left(\frac{\kappa_0-\kappa_{m,i}}{\kappa_0+\kappa_{m,i}}\right)\left(1-\ee{-2
d_i\kappa_{m,i}}\right)}{1-\left(\frac{\kappa_0-\kappa_{m,i}}{\kappa_0+\kappa_{m,i}}\right)^{2}\ee{-2d_i\kappa_{m,i}}}\,
,\quad
 t_i&=\frac{\left[1-\left(\frac{\kappa_0-\kappa_{m,i}}{\kappa_0+\kappa_{m,i}}\right)^{2}\right]\ee{-d_i\kappa_{m,i}}}{
1-\left(\frac{\kappa_0-\kappa_{m,i}}{\kappa_0+\kappa_{m,i}}\right)^{2}\ee{-2d_i\kappa_{m,i}}}, 
\end{eqnarray}
where $\kappa_{m,i}=\sqrt{k^2-\varepsilon_i\omega^2}$ and $i=1,2$. Anticipating that we 
will investigate phenomena in the low THz regime, we approximate the material permittivities
by their static value. Further, in our dimensionless units, the slabs' thicknesses $d_i$ 
undergo the replacement $2\gap d_i/\hbar c\to d_i$.

The above expressions describe several different setups where graphene is combined with 
a dielectric environment and in the following we focus on four specific configurations: 
First, we investigate a setup where a single graphene layer is embedded in an infinitely 
thick dielectric material ($d_i\rightarrow\infty$, $\varepsilon_i\equiv\varepsilon_m$). 
Then, we consider the case of a  where a graphene layer is 
placed between two identical dielectric slabs with finite thickness ($d_i=d$, 
$\varepsilon_i\equiv\varepsilon_m$). Next, we generalize the above ``sandwich'' configuration 
to an asymmetric setup where a graphene layer is deposited first on an infinite dielectric 
substrate ($d_1\to \infty$, $d_2\rightarrow\infty$, $\varepsilon_1=1$, $\varepsilon_2\equiv\varepsilon_m$) 
and, finally, we consider a graphene layer on a dielectric slab with finite thickness 
($d_1\to \infty$, $d_2\equiv d$). 
For all configurations we will determine the structures' resonances in the response to 
TE-polarized electromagnetic radiation. Mathematically, this corresponds to investigating 
the condition 
\begin{eqnarray}
\label{eq:poles} 
 \frac{1}{r}=0&\qquad\Leftrightarrow\qquad (r_1r_g-1)(r_2r_g-1)=r_1r_2t_g^2~.
\end{eqnarray}
This provides the starting point of all analyses in the subsequent sections.

\begin{figure}[t!]
\centering
  \includegraphics{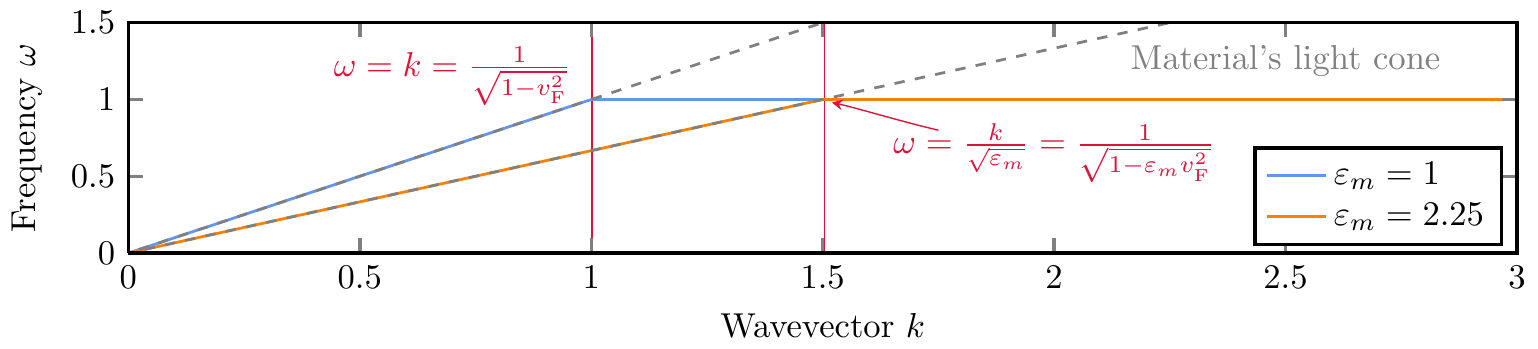}
\caption{Dispersion relation of the TE plasmon for a freely suspended graphene 
         layer ($\varepsilon_m = 1$) and when embedded in an infinite dielectric 
         with permittivity $\varepsilon_m = 2.25$.
         }
\label{fig:TEplasmonMode_backgroundEps}
\end{figure}

\section{Graphene embedded in an infinitely extended dielectric}

When a graphene layer is embedded in an infinitely extended dielectric Eq.~\eref{eq:poles} is equivalent 
to $\alpha\Phi(y)+2\kappa_m=0$. For freely suspended graphene ($\varepsilon_m=1$), the numerical solutions 
of this equation have already been analyzed in Ref.~
\cite{Bordag_2014,Bordag_2015}.
Here, we rather present the dispersion relation in terms of a parametric expression. Setting $y=\tanh{(q)}$, 
where $q$ is the external parameter, we obtain that the TE resonance is described by
\numparts
\begin{eqnarray}
 \label{eq:TEPlasmon_vacuum_k}
&&k[q]=\frac{1}{\sqrt{1-\varepsilon_m\vF^2}}\sqrt{\varepsilon_m\tanh{(q)}^2+\alpha^2\left(q\frac{\tanh{(q)}
^2+1}{\tanh{ (q) } } -1\right)^2 } 
\\
 \label{eq:TEPlasmon_vacuum_w}
&&\omega[q]=\frac{1}{\sqrt{1-\varepsilon_m\vF^2}}\sqrt{\tanh{(q)}^2+\alpha^2\vF^2\left(q\frac{\tanh{(q)}
^2+1}{\tanh{(q) } } -1\right)^2}~.
\end{eqnarray}
\endnumparts
In Fig.~\ref{fig:TEplasmonMode_backgroundEps}, we depict the dispersion relation \eref{eq:TEPlasmon_vacuum_k},
and \eref{eq:TEPlasmon_vacuum_w} 
for $\varepsilon_m = 2.25$ and for vacuum $\varepsilon_m = 1$. The latter curve agrees with the corresponding
numerical results~
\cite{Bordag_2014}. 
The complete TE-plasmon dispersion relation lies in the embedding medium's evanescent region, i.e., below 
the embedding medium's light cone 
($\omega<k/\sqrt{\varepsilon_{m}}$).
We can clearly distinguish two regimes: For low frequencies, $\omega < 1$ (or, equivalently, 
$k < \sqrt{\varepsilon_{m}}$), the TE-plasmon dispersion relation is close to the light cone 
$\omega = k/\sqrt{\varepsilon_{m}}$ and can be well approximated by 
$\omega[k]=\beta(k)\, k$, where
\begin{eqnarray}
\label{eq:plasmonSmaller}
\beta(k)&\approx\frac{\sqrt{1-\frac{16}{9}\alpha^2\left(\frac{1-\varepsilon_m\vF^2}{\varepsilon_m}
\right)^2k^2 } }{\sqrt{\varepsilon_m} }\lesssim
\frac{1}{\sqrt{\varepsilon_m}},
\end{eqnarray}
For high frequencies, $\omega \gtrsim 1$ (or, equivalently $k > \sqrt{\varepsilon_{m}}$), the TE-plasmon
dispersion relation is practically independent of the embedding dielectrics' properties 
($q\gtrsim1$)
\begin{eqnarray}
\label{eq:plasmonLarger}
&&k[q]\approx\sqrt{\frac{\varepsilon_m(2\tanh{(q)}-1)+\alpha^2(2q-1)^2}{1-\varepsilon_m\vF^2}},\\
&&\omega[q]\approx\sqrt{
\tanh{(q)}^2+\vF^2k^2}~.\nonumber
\end{eqnarray}
so that for these frequencies the dispersion relation becomes very flat and takes on an almost 
constant value for $\vF\ll1$.
The transition between these two rather distinct regimes of the dispersion relations occurs
near $\omega \approx 1$ (or, equivalently, $k \approx \sqrt{\varepsilon_{m}}$). Due to the
small value of the fine structure constant $\alpha$, this transition range is rather small,
so that the transition appears to be rather abrupt. However, formally studying the behavior 
of the dispersion relation for larger values of $\alpha$, reveals that this apparent ``kink'' 
is actually a smooth crossover
\cite{Bordag_2014,Bordag_2015}.
In the following, we will utilize this formal trick of considering larger values of $\alpha$
for better representing this feature
\cite{Bordag_2014,Bordag_2015}. 

\begin{figure}[t!]
\centering
  \includegraphics[width=\linewidth]{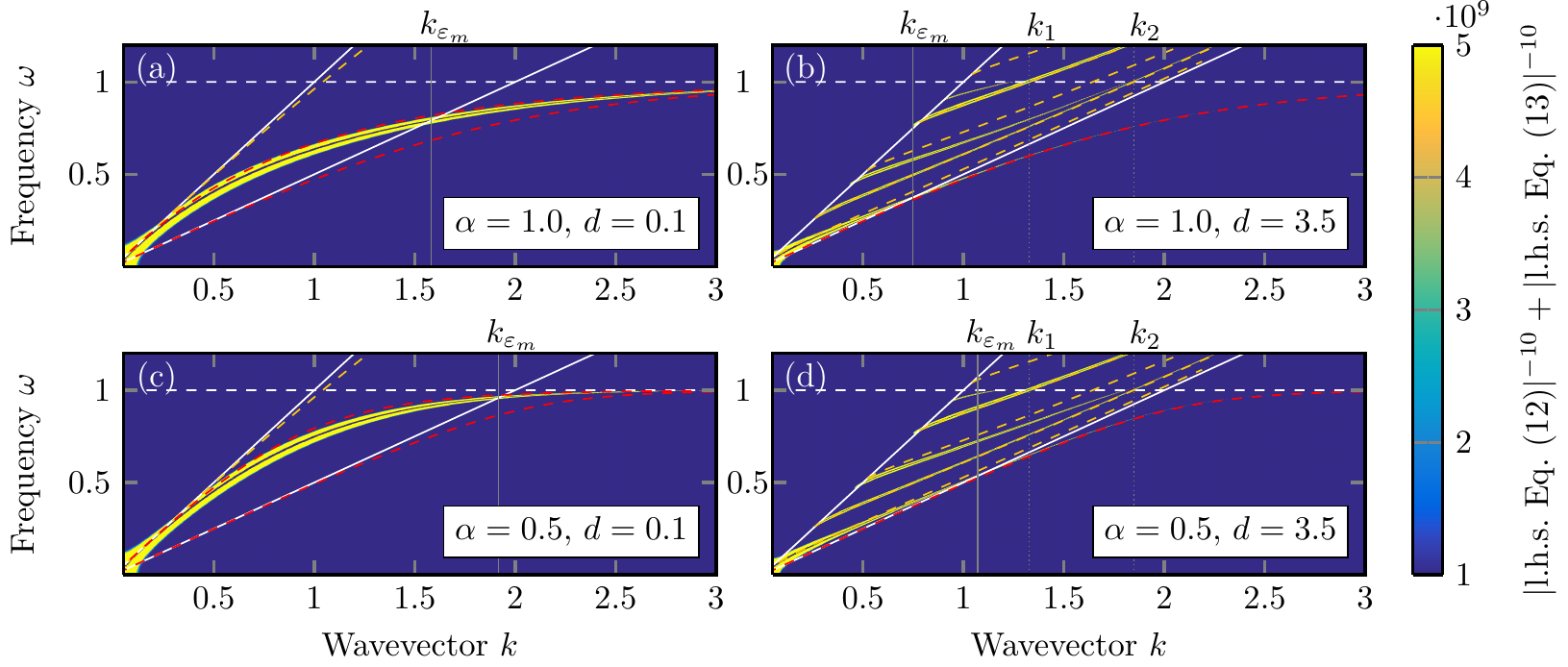}
\caption{TE-plasmon resonances for a slab-graphene-slab setup with $\varepsilon_{m}=4.0$ and 
         different thickness $d$ of the dielectric slabs. For representation purposes we use 
         values $\alpha=1,1/2$ which artifically enlarge the transition region
         \cite{Bordag_2014,Bordag_2015}. 
         Color-coded is the behavior of the inverse absolute value of the l.h.s. of 
         Eq.~\eref{eq:plasmonEvenDispersionSlab} and the equivalent for the solution $r\TE
         _m=-1$. The orange dashed lines describe the modes of a single dielectric slab 
         (with $d'=2d$ and without graphene). The dashed red line represent the TE-plasmon 
         resonances (see Eqs.~\eref{eq:TEPlasmon_vacuum_k} and \eref{eq:TEPlasmon_vacuum_w}) for 
         graphene suspended in vacuum (upper curve) and embedded in an infinite dielectric with 
         $\varepsilon_m=4.0$ (lower curve). 
         The white solid lines represent vacuum and material light cones. The dashed line marks 
         the pair-creation threshold with $\omega=\sqrt{1+\vF^2k^2}$ (see text for 
         further details).
         }
\label{fig:modes_grapheneSlab}
\end{figure}

\section{A graphene layer between two identical dielectric slabs with finite thickness}

The physics becomes considerably richer when a graphene layer is sandwiched between two identical 
dielectric slabs of finite thickness $d$. For this configuration, Eq.~\eref{eq:poles} leads
to two different sets of modes. The first set is given by the solutions of 
\begin{eqnarray}
\label{eq:plasmonOddDispersionSlab}
 r_{m}\TE+1&=0
\end{eqnarray}
which corresponds to waveguide modes with odd symmetry for a slab of thickness $d'=2d$ (see
Fig.~\ref{fig:modes_grapheneSlab}). Since these modes have zero field at the position of the 
graphene sheet, they are not affected by its presence. Therefore, while we retain these unaltered modes,
we concentrate in the following discussion mainly on the altered modes. Equation~\eref{eq:poles} provides, in fact, a
second (the even) set of modes which are the solution of
\begin{eqnarray}
\label{eq:plasmonEvenDispersionSlab}
\alpha\Phi(y)&\left[\kappa_0+\kappa_m-\ee{-2d\kappa_m}\left(\kappa_0-\kappa_m\right)\right]
\nonumber\\&+2\kappa_m\left[\kappa_m+\kappa_0+\ee{-2d\kappa_m}\left(\kappa_0-\kappa_m\right)\right] &= 0\,.
\end{eqnarray}
This can be rewritten in the more convenient form
\begin{eqnarray}
 \label{eq:conditionThicknessSlab}
 d&=\frac{1}{2\kappa_m}\ln{\left[\frac{\left(\kappa_0-\kappa_m\right)\left(\alpha\Phi(y)-2\kappa_m\right)}{
\left(\kappa_0+\kappa_m\right)\left(\alpha\Phi(y)+2\kappa_m\right)}\right]}\,.
\end{eqnarray}
Owing to the fact that $d$ is a real number, Eq.~\eref{eq:conditionThicknessSlab} allows for 
real valued solutions only in certain regions of the $(k,\omega)$-plane. Indeed, when $k<\omega$ (propagating 
waves in vacuum) neither damped nor undamped waveguide modes exist ($d$ is always complex since
$\kappa_0-\kappa_m<0$). Additionally, for $y>1$ (the single particle excitation region of graphene above the
pair-creation threshold), the r.h.s. of the Eq.~\eref{eq:conditionThicknessSlab} necessarily has an nonzero imaginary
part and therefore only damped modes can exist (see below).

\begin{figure}[t!]
\centering
  \includegraphics{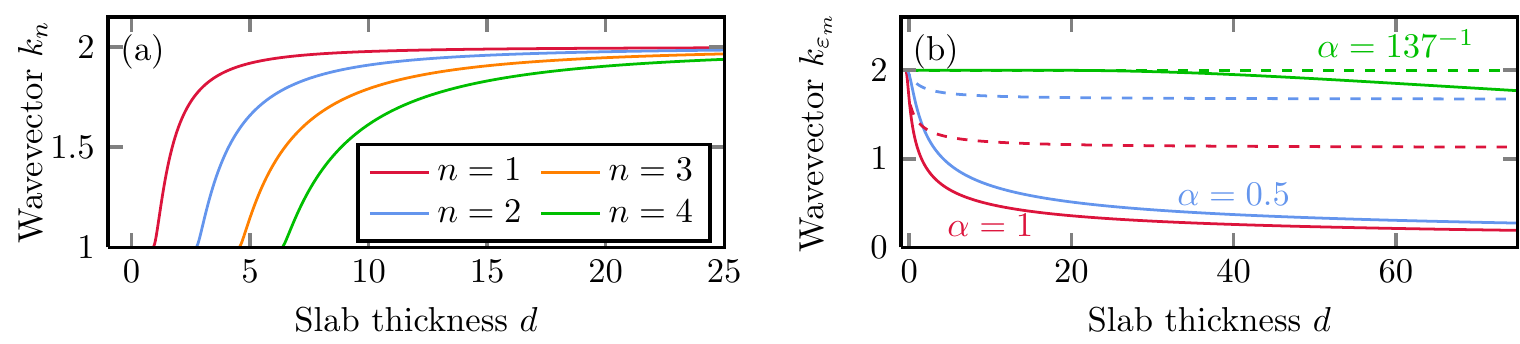}
\caption{Panel (a): Thickness dependence of the cutoff wavevectors $k_{n}[d]$ for $n=1,2,3,4$ that 
                    determine the behavior of waveguiding in graphene-dielectric-slab systems. 
         Panel (b): Thickness dependence of the wavevector for which the TE-plasmon resonance 
                    crosses the dielectric material's light cone. We consider different values
                    of $\alpha$ and show the results for graphene on a slab (dashed) and graphene 
                    between two slabs (solid) in order to connect to the results of Figs.~\ref{fig:modes_grapheneSlab}
                    and \ref{fig:modes_grapheneFiniteSubstrate}. See text for further details.
        }
\label{fig:cutOffFrequencies_grapheneSlab}
\end{figure}

In the remaining regions, Eq.~\eref{eq:conditionThicknessSlab} exhibits two sets of physically 
distinct solutions.
The first set of modes is located in the region $\omega<k<\omega\sqrt{\varepsilon_m}$ with $y\leqslant1$ 
and corresponds to guided, real frequency modes propagating in the slab, i.e., modes that exhibit an evanescent
orthogonal wavevector 
in vacuum. The existence of these solutions can be understood by noticing that in this region 
$k_{m}^{\perp}=\ii \kappa_{m}$ and $-\alpha\Phi(y)$ are positive numbers. 
Upon rewriting Eq.~\eref{eq:conditionThicknessSlab} we, therefore, obtain
\begin{eqnarray}
 \label{eq:dConditionWaveguide}
 d&=\frac{\atan{\frac{2k_{m}^{\perp}}{-\alpha\Phi(y)}}-\atan{\frac{2k_{m}^{\perp}}{2\kappa_0}
}+\pi n}{k_{m}^{\perp}}>0,\,
\end{eqnarray}
where $n=1,2,3,\dots$ (see below for the case $n=0$). Each value of $n$ corresponds to a different 
waveguide mode. Interestingly and in contrast to ordinary guided modes in a dielectric slab, the 
incorporation of a graphene layer sets a cutoff to the dispersion relations for these lossless modes on the 
line $y=1$. We display these modes -- together with the undisturbed, odd modes ($r\TE_m=-1$) -- in
Fig.~\ref{fig:modes_grapheneSlab} where, for clarity of the
presentation, we have used values $\alpha=1,\,1/2$ for the fine structure constant.
The actual values of the cutoff
wavevectors $k_{n}$, for which the waveguide modes cross the cutoff line, depend on the value of $n$, on the dielectric
function and the thickness of the slab. For each $n$ the function $k_{n}[d]$ can be described in terms of the following
parametrization
\numparts
\begin{eqnarray}
\label{eq:dvalues}
 d[k_{m}^{\perp}]&=\frac{2\pi n-\atan{\frac{\sqrt{1-\varepsilon_m\vF^2}k_{m}^{\perp}}{\sqrt{
\left(\varepsilon_m-1\right)-\left(1-\varepsilon_m\vF^2\right)\left(k_{m}^{\perp}\right)^2} }} } { k_{m}^{\perp}}\,,
\\
k_{n}[k_{m}^{\perp}]&=\sqrt{\frac{\varepsilon_m-\left(k_{m}^{\perp}\right)^2}{1-\varepsilon_m\vF^2}},
\label{eq:knvalues}
\end{eqnarray}
\endnumparts
where $k_{m}^{\perp}>0$ [see Fig.~\ref{fig:cutOffFrequencies_grapheneSlab}(a)]. Interestingly, the
$k_n[d]$ given in Eq.~\eref{eq:dvalues} and~\eref{eq:knvalues} are equal to the condition for the values of the
wavevector at which the undisturbed odd waveguide modes cross the value $y=1$. From Fig.~\ref{fig:modes_grapheneSlab} we
can see that, in the region $y>1$, the modes modified by the presence of graphene run between the even and odd modes of
the pure dielectric waveguide.

\begin{figure}[t!]
\centering
  \includegraphics{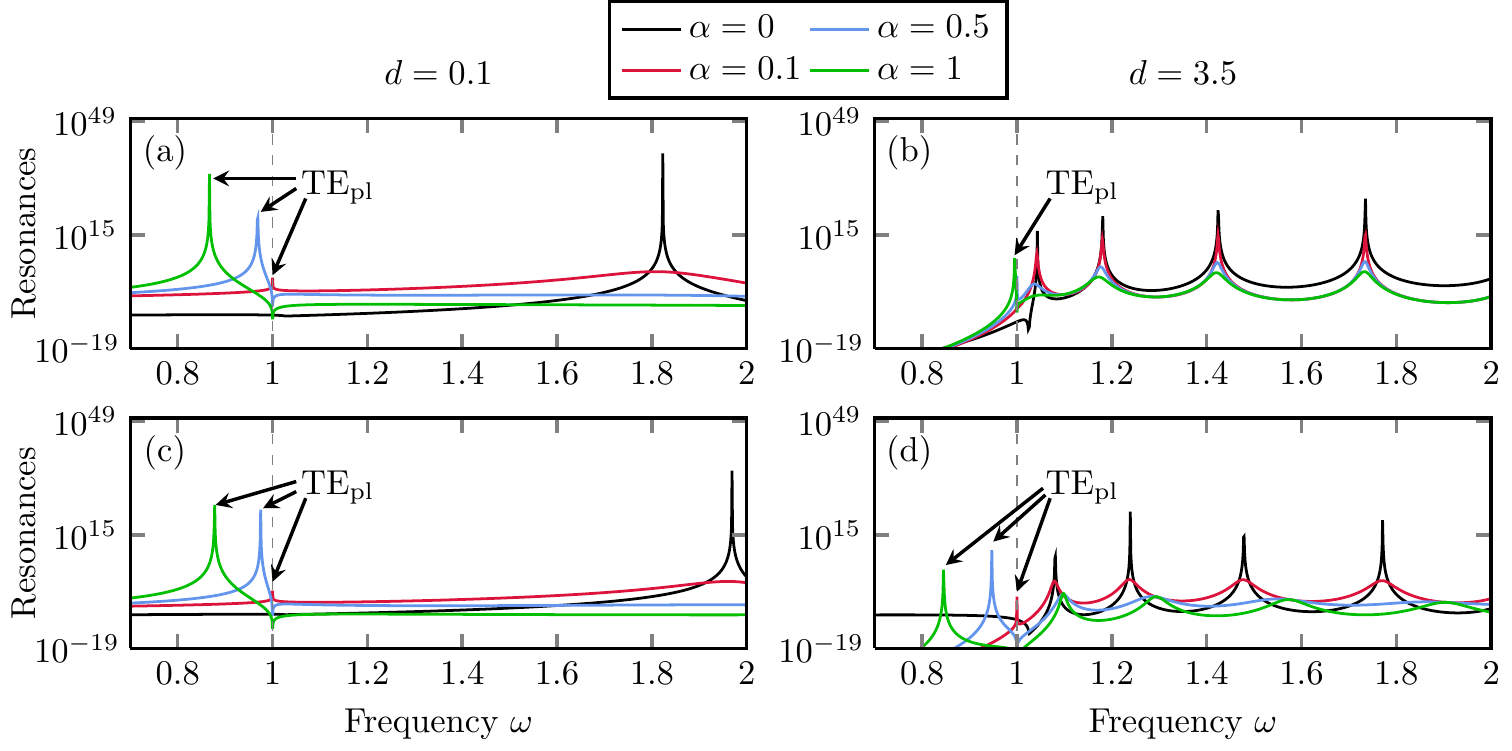}
\caption{The Figure above shows the (damped) modes for $k_0=2.05$ and two different dielectric ($\varepsilon_m=4$)
         slab thicknesses. The TE plasmonic resonance are denoted by TE$_{\mathrm{pl}}$ and can be found
         in the region $y<1$.
         Panel (a,b): Solutions for graphene embedded between two slabs of thickness $d$. Depicted as
         resonance is the quantity $(|$l.h.s of Eq.~\eref{eq:plasmonOddDispersionSlab}$|^{-10} + $l.h.s of
          Eq.~\eref{eq:plasmonEvenDispersionSlab}$|^{-10})$. $\alpha=0$ 
         shows only the even solutions of a waveguide without graphene of thickness $d'=2d$
         [Eq.~\eref{eq:plasmonEvenDispersionSlab} with $\alpha=0$]. Only these even modes that are then 
         altered when including graphene. 
         Panel (c,d): Solutions for graphene embedded on top of a dielectric slab, where the plots show
         the quantity $|$l.h.s of Eq.~\eref{eq:plasmonDispersionFiniteSubstrate}$|^{-10}$. $\alpha=0$ shows 
         the solution for a purely dielectric waveguide of thickness $d'=d$.
        }
\label{fig:damping_grapheneSlab}
\end{figure}

From the previous reasoning it follows that in the region $y>1$ only damped, even modes are allowed. Indeed, when
graphene's pair-creation threshold is reached, a decay channel is introduced in the optical response of the system.
Describing the slab-graphene-slab setup by an effective dielectric medium for the even modes and assuming
that this effective medium $\varepsilon_{\mathrm{eff}}=\varepsilon+\varepsilon^{(1)}$ with
$\varepsilon^{(1)}\ll\varepsilon$, we find that $\varepsilon^{(1)}\propto\alpha$ and thus also the damping is
proportional to $\alpha$. We remark that in Fig.~\ref{fig:modes_grapheneSlab} the even modes cannot be discerned
since for large $\alpha$ the damping is very dominant. However, the modes and the corresponding damping become visible
in Fig.~\ref{fig:damping_grapheneSlab}(a,b) where we show cuts through Fig.~\ref{fig:modes_grapheneSlab} at $k_0=2.05$
while varying $\alpha$.

Returning to Eq.~\eref{eq:dConditionWaveguide}, we note that a second solution for $n=0$ is possible 
if $2\kappa_0+\alpha\Phi(y)>0$. This corresponds to a mode with a dispersion relation always located 
below that of the vacuum TE-plasmon resonance (see Fig.~\ref{fig:modes_grapheneSlab}). 
Curiously, this dispersion relation describes a field which is propagating along the direction 
orthogonal to the graphene layer for $0<k<k_{\varepsilon_{m}}$ and evanescent for $k>k_{\varepsilon_{m}}$ 
\cite{Intravaia_2005,Intravaia_2007}. 
As before, the value of $k_{\varepsilon_{m}}$ depends on the systems parameters and can be found by 
solving
\begin{eqnarray}
  d &= \lim_{\kappa_m\to0}\left\{\frac{1}{2\kappa_m}\ln{\left[\frac{
\left(\kappa_0-\kappa_m\right)\left(\alpha\Phi(\omega,k)-2\kappa_m\right)}{
\left(\kappa_0+\kappa_m\right)\left(\alpha\Phi(\omega,k)+2\kappa_m\right)}\right]}\right\}\nonumber\\&=-\frac{
\alpha\Phi(y)+2\kappa_
0} { \alpha\kappa_0\Phi(y)}\,.
\end{eqnarray}
This leads to the parametric representation
\numparts 
\begin{eqnarray}
  k_{\varepsilon_m}[q] &= \sqrt{\frac{\varepsilon_m}{1-\varepsilon_m\vF^2}}\tanh{(q)}\,,\\
 d[q] &=
-\frac{\alpha\Phi[\tanh{(q)}]+2\sqrt{\frac{\varepsilon_m-1}{1-\varepsilon_m\vF^2}}\tanh{(q)}}{\alpha\sqrt{\frac{
\varepsilon_m-1}{1-\varepsilon_m\vF^2}}\tanh{(q)}\Phi[\tanh{(q)}]}
\end{eqnarray}
\endnumparts
which we depict in Fig.~\ref{fig:cutOffFrequencies_grapheneSlab}(b).  
For $k\ll k_{\varepsilon_{m}}$, this dispersion relation starts along the vacuum light cone $\omega\lesssim k$, 
while in regions where the mode becomes evanescent ($k>k_{\varepsilon_{m}}$) the condition
$2\kappa_m+\alpha\Phi(y)\leq0$ 
must be satisfied. This implies that the dispersion relation of this mode lies above the TE-plasmon
resonance of graphene embedded in an infinite material with dielectric permittivity $\varepsilon_m$ 
(see Fig.~\ref{fig:modes_grapheneSlab}). Similar to this latter case, in the limit $k\to \infty$ the 
mode becomes independent of the material properties, approaching $\omega=\sqrt{1+\vF^2k^2}$ from below.

The different physics associated with the two sets of modes also has an impact on their characteristics 
for increasing the slab's thickness. For $d\to \infty$ the number of waveguide modes increases but their 
dispersion relations remain confined in the region $\omega<k<\omega\sqrt{\varepsilon_m}$. 
The modes become denser and denser tending to the scattering waves of the infinite system (mathematically 
represented in the equations by a branch cut). In contrast, the singularity corresponding to the $n=0$ 
solution remains isolated and actually approaches the dispersion relation given in Eqs.
\eref{eq:TEPlasmon_vacuum_k} and \eref{eq:TEPlasmon_vacuum_w}.

\section{Graphene on a substrate}

\begin{figure}[t!]
\centering
  \includegraphics{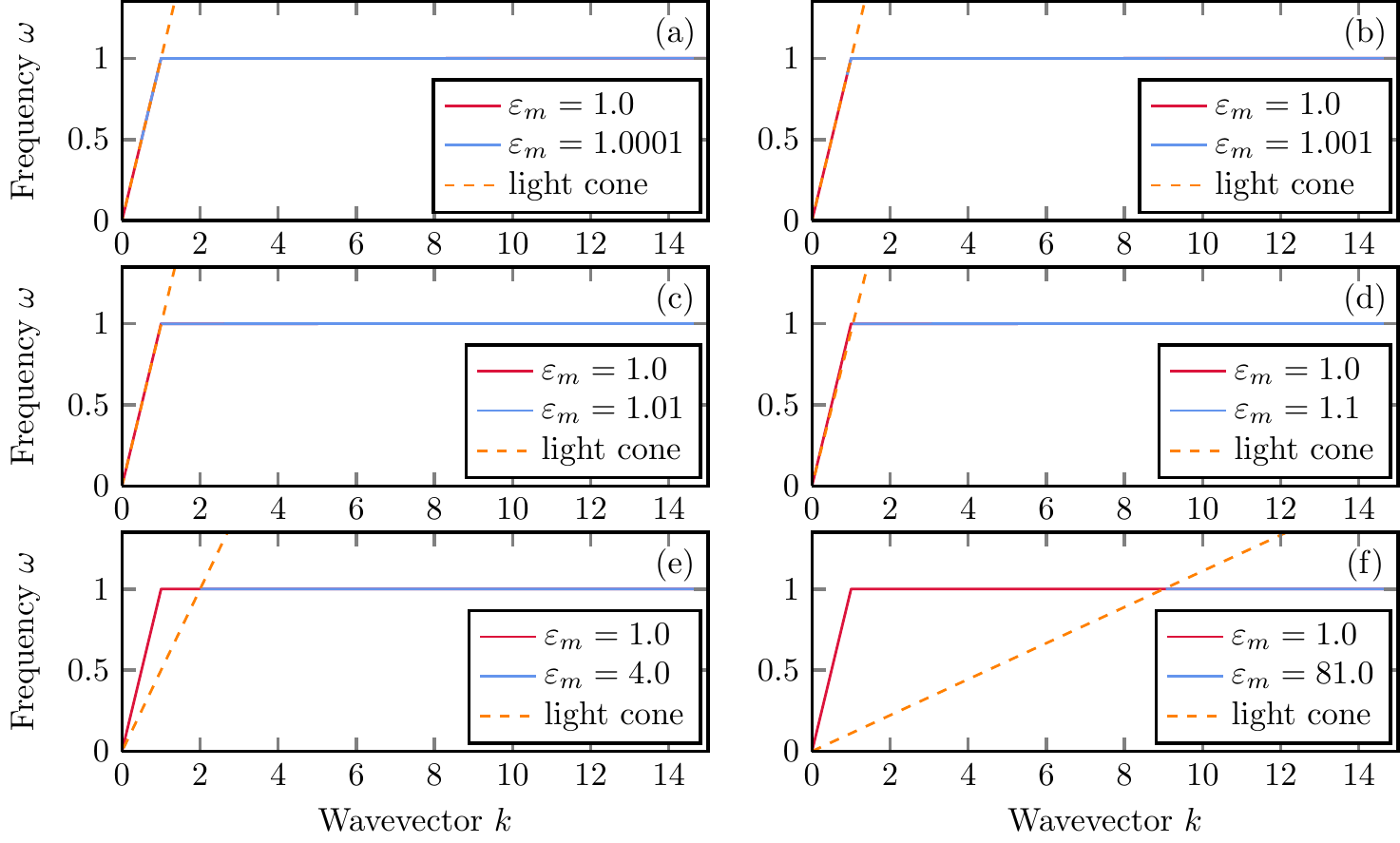}
\caption{Dispersion relation of the TE surface plasmon in a graphene layer deposited on 
         a bulk substrate with dielectric constant $\varepsilon_m$. In all subfigures,
         the substrate material's light cone is indicated by the dashed orange line. 
         The red line represents the TE-plasmon dispersion relation for a graphene
         layer in vacuum.}
\label{fig:TEplasmonMode_substrate}
\end{figure}
In most experimental setups graphene is deposited on dielectric substrates, i.e., in
terms of our above analysis $\varepsilon_1=1$ and $\varepsilon_2=\varepsilon_m$ with 
different thicknesses. 
In order to understand how the TE-plasmon resonances behave in this case, it is convenient 
to first consider the bulk limit case ($d_{1(2)}\rightarrow\infty$). Using $r_1(\kappa_{m,1}=\kappa_0)=0$ and
$r_2\equiv r_m(d\to\infty)=\frac{\kappa_0-\kappa_m}{\kappa_0+\kappa_m}$, Eq.~\eref{eq:poles} 
leads to
\begin{eqnarray}
\label{eq:conditionPlasmonTwoMedia}
 r_mr_g-1=0\quad\Leftrightarrow\quad\alpha\Phi(y)+\kappa_0+\kappa_m&=0~.
\end{eqnarray}
This has the parametric solution
\numparts
\begin{eqnarray}
\label{eq:kubstrate}
k^2[y] &=\frac{\omega[y]^2-y^2}{\vF^2}\\
\label{eq:wubstrate}
\omega^2[y] &=\frac{4
y^2+\vF^2\alpha^2\Phi(y)^2
}{\left[2-(\varepsilon_m+1)\vF^2\right]\left\{1+\sqrt{1-\left(\frac{\left(\varepsilon_m-1\right)\vF}{2
-(\varepsilon_m+1)\vF^2}\right)^2\frac{4y^2+\vF^2\alpha^2\Phi(y)^2}{\alpha^2\Phi(y)^2}}\right\}}~,
\end{eqnarray}
\endnumparts
supplemented by the additional condition $\alpha^2\Phi(y)^2\geqslant\kappa_0^2+\kappa_m^2$. 
In Fig.~\ref{fig:TEplasmonMode_substrate}, we display the corresponding dispersion relation and 
note that for $k>\omega \sqrt{\epsilon_{m}}$ it is rather similar to the dispersion relation 
for a gapped graphene layer in vacuum . Effectively, the material light cone cuts the dispersion
relation for the TE-plasmon resonance and `removes' the values for
$k<\sqrt{\varepsilon_{m}/(1-\varepsilon_{m}\vF^{2})}$.
For the parameters $\alpha=137^{-1}$ and $\vF=300^{-1}$, already a substrate with a $\varepsilon_m > 1.1$
(see panel (d) in Fig.\ref{fig:TEplasmonMode_substrate})
leads to dispersion relation that does not anymore depend on the properties of the dielectric. Despite 
this, a solution exists also for high values of the permittivity as long as $\varepsilon_m<\vF^{-2}$.

\begin{figure}[t!]
\centering
  \includegraphics[width=\linewidth]{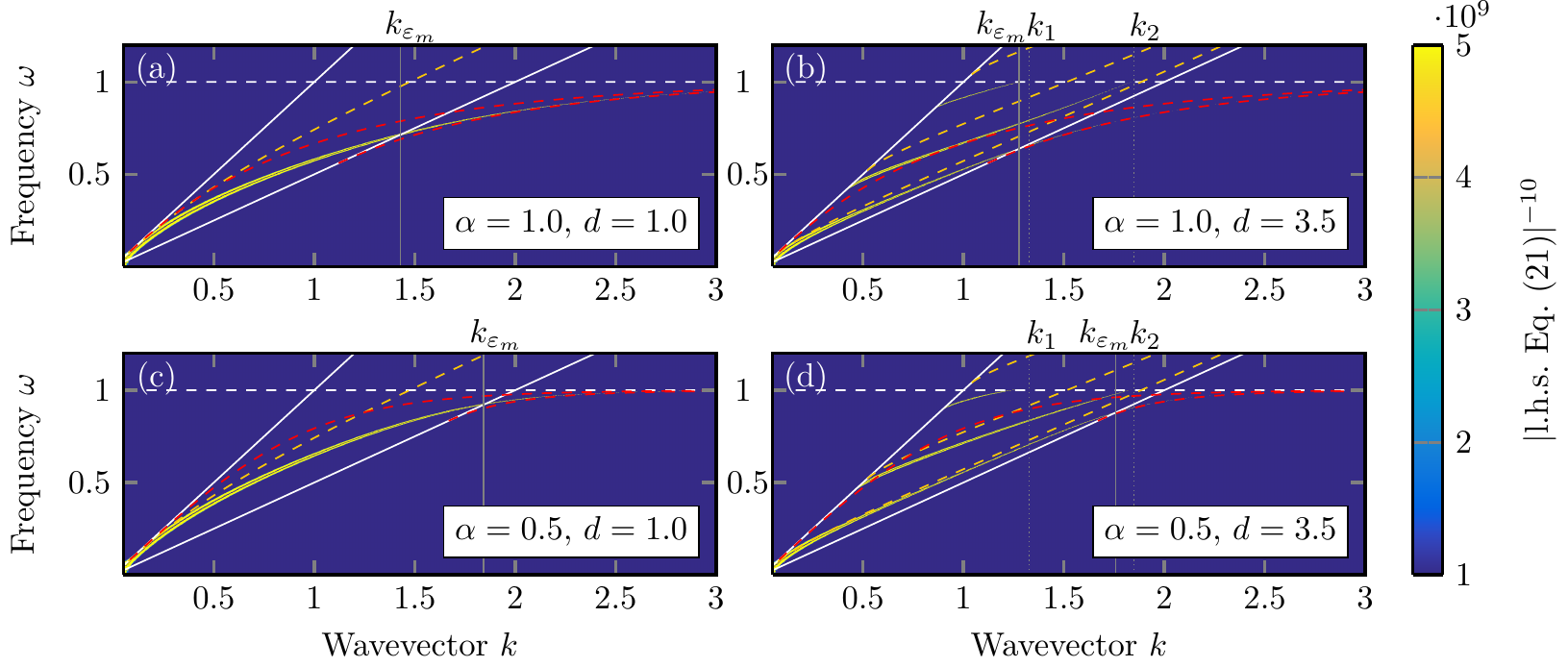}
\caption{TE-plasmon resonances for a graphene layer on a dielectric substrate with finite 
         thickness. The slab's waveguide modes are indicated by the dashed orange lines. 
         A TE-plasmon resonance that crosses the material's light cone is discernible. 
         The red dashed lines represent the solution of the TE-plasmon mode for a graphene 
         layer in vacuum (upper row) and a graphene layer deposited on an infinitely extended 
         substrate with dielectric function $\varepsilon_m$ (lower row).
         The remaining parameters are the same as in Fig.~\ref{fig:modes_grapheneSlab}.}
\label{fig:modes_grapheneFiniteSubstrate}
\end{figure}

More interesting is the case where the substrate has a finite thickness $d$. Similar to 
the two-slab configuration, the resonance condition 
\begin{eqnarray}
\label{eq:plasmonDispersionFiniteSubstrate}
-\alpha\Phi(y)&\left[(\kappa_0+\kappa_m)-(\kappa_0-\kappa_m)\ee{-2d\kappa_m}\right]\nonumber\\
&-(\kappa_0+\kappa_m)^2+(\kappa_0-\kappa_m)^2\ee{-2d\kappa_m}=0\,, 
\end{eqnarray}
can be recast as
\begin{eqnarray}
 \label{eq:conditionThicknessFiniteSubstrate}
d=\frac{1}{2\kappa_m}\ln{\left(\frac{\left(\kappa_0-\kappa_m\right)\left(\alpha\Phi(y)+\kappa_0-\kappa_m\right)}{
\left(\kappa_0+\kappa_m\right)\left(\alpha\Phi(y)+\kappa_0+\kappa_m\right)}\right)}\,.
\end{eqnarray}
Due to the asymmetry of this setup, opposite to Eq.~\eref{eq:plasmonEvenDispersionSlab} all waveguide modes are
altered by the presence of graphene. The analysis of Eq.~\eref{eq:conditionThicknessFiniteSubstrate} proceeds along
lines that are similar to those discussed in the previous section. Also in this case, two 
sets of modes are possible (see Fig.~\ref{fig:modes_grapheneFiniteSubstrate}). 
We obtain a TE-plasmon resonance which exists for all $k$, crossing the material light 
cone for $k=k_{\varepsilon_m}$.
In this latter case the $k_{\varepsilon_m}$ has the following dependence on the substrate 
thickness
\numparts
\begin{eqnarray}
  k_{\varepsilon_m}[q] &= \sqrt{\frac{\varepsilon_m}{1-\varepsilon_m\vF^2}}\tanh{(q)}\,, \\
 d[q] &=
-\frac{\alpha\Phi[\tanh{(q)}]+2\sqrt{\frac{\varepsilon_m-1}{1-\varepsilon_m\vF^2}}\tanh{(q)}}{\left\{\alpha
\Phi[\tanh{(q)}] +\sqrt{\frac{
\varepsilon_m-1}{1-\varepsilon_m\vF^2}}\tanh{(q)}\right\}\sqrt{\frac{
\varepsilon_m-1}{1-\varepsilon_m\vF^2}}\tanh{(q)} }.
\end{eqnarray}
\endnumparts
In Fig.~\ref{fig:cutOffFrequencies_grapheneSlab}, we represent this solution by dashed lines. 
The dispersion relation describing this resonance lies between the symmetric vacuum expression 
given in Eqs.~\eref{eq:TEPlasmon_vacuum_k} and \eref{eq:TEPlasmon_vacuum_w} and the asymmetric 
bulk dispersion relation of Eqs.~\eref{eq:kubstrate} and \eref{eq:wubstrate} (red dashed lines 
in Fig.~\ref{fig:modes_grapheneFiniteSubstrate}), rapidly approaching this last curve for 
$k>k_{\varepsilon_m}$. 
For $k\to 0$ the dispersion relation of the TE-plasmon follows the vacuum light cone and shows
the same asymptotic behavior as the lowest-frequency solution of
\begin{eqnarray}
\label{eq:waveguideSlab}
 (\kappa_0+\kappa_m)^2-(\kappa_0-\kappa_m)^2\ee{-2d\kappa_m}&=0\,,
\end{eqnarray}
the solutions of which represent ordinary waveguiding modes in simple dielectric slabs. In
Figs.~\ref{fig:modes_grapheneFiniteSubstrate}, we observe that indeed the TE-plasmon dispersion 
relation approaches the orange dashed line that describes the zeroth waveguiding mode of a
single dielectric slab.

As in the double-slab configuration discussed in the previous section, we find here that undamped waveguide 
modes only exist in the regions $\omega<k<\omega\sqrt{\varepsilon_m}$ and $y\le 1$. Due to pair-creation,
analogously to the slab-graphene-slab case, damped waveguide modes exist for $y>1$ [cf. 
Fig.~\ref{fig:damping_grapheneSlab}(c,d)]. Here, the damping also increases with increasing $\alpha$. Once
again, 
the graphene layer imposes a cutoff on the real solutions of Eq.~\eref{eq:waveguideSlab} and the 
undamped modes end on $y=1$ for values of $k_{n}$ given as in the symmetric two-slab 
case by Eqs.~\eref{eq:dvalues} and \eref{eq:knvalues}.

\section{Conclusions}

We have investigated the electromagnetic characteristics of different graphene-dielectric 
structures and have demonstrated that, when a gap appears in the graphene's band structure,
TE-plasmon resonances exist for large values of the dielectric permittivity.
Depending on the actual realization, the characteristics of the corresponding dispersion 
relations are quite different, in certain cases describing waves that change their nature 
from propagating to evanescent as a function of the wavevector. When graphene is deposited 
on a bulk substrate only the evanescent part of the dispersion relation remains as an 
isolated singularity, while the propagating part merges in the continuum of scattering 
states in the semi-infinte medium.
We have further shown, that when graphene is in contact with one or two slabs of finite 
thickness, the slab's real waveguiding modes exhibit a cutoff frequency whose value corresponds 
to the energy threshold for pair-creation in graphene (single particle excitation threshold). Modes with frequency
above this cutoff are damped since the pair-creation constitutes a loss channel.
Quantum properties of graphene, specifically the pair-creation, thus affect both, the TE-plasmon 
properties and the waveguiding modes introducing additional losses in the electromagnetic
response. Because these effects are related to the existence of a band gap, they do not 
appear in recent studies that address waveguiding modes for gapless graphene-dielectric 
structures~
\cite{Hanson_2008_JAP2} (see also very recent studies for metal-graphene setups~
\cite{Khromova_2014,Ralevic_2014,Luo_2015,Shkerdin_2015,Degli_2015}). If we were to include additional loss
channels into the description of graphene such as a finite relaxation rate for the carriers in
graphene, these losses would increase.

The above results are of interest for several experimentally relevant scenarios such as 
atom-chip systems. In these systems, a cloud of cold or ultra-cold atoms are magnetically 
trapped near surfaces~
\cite{Fortagh_2007,Folman_2002}. 
As TE-polarized fields have a large magnetic component, the TE-plasmons and all further TE waveguiding modes
associated with 
graphene-dielectric structures may profoundly affect the dynamics of magnetic emitters as they provide
additional decay channels. The thorough understanding of these modes is thus key to the correct description and
characterization of the decay dynamics of such emitters. 
Finally, the opportunity 
to damp and thus filter certain waveguiding modes by including a graphene layer with tunable 
band gap~\cite{Jung_2015} may open novel paths for technological applications, notably in the area of nano-photonic
systems.

\section{Acknowledgements}
We acknowledge support by the Deutsche Forschungsgemeinschaft (DFG) through the Collaborative Research 
Center (CRC) 951 ``Hybrid Inoganic/Organic Systems for Optoelectronics (HIOS)" within project B10.
FI further acknowledges financial support from the EU through the Career Integration Grant (CIG) 
No. 631571 and support by the DFG through the DIP program (FO 703/2-1).\\

\providecommand{\newblock}{}

\end{document}